\documentstyle[aps,prb,multicol,psfig]{revtex} 
\newif\ifmynarrow
\mynarrowfalse
\renewcommand{\narrowtext}{%
  \ifmynarrow\hspace*{\fill}\raisebox{-1ex}[0pt][0pt]{%
    \rule{0.3pt}{1ex}%
    \rule[1ex]{20.5pc}{0.3pt}}\fi \mynarrowtrue
  \vspace{-1.0ex}%
  \begin{multicols}{2}%
  \par\global\columnwidth20.5pc
  \global\hsize\columnwidth\global\linewidth\columnwidth
  \global\displaywidth\columnwidth}
\renewcommand{\widetext}{%
  \end{multicols}%
  \vspace{-2.5ex}%
  \noindent\raisebox{1ex}[0pt][0pt]{%
    \rule{20.5pc}{0.3pt}%
    \rule{0.3pt}{1ex}}%
  \par\global\columnwidth42.5pc
  \global\hsize\columnwidth\global\linewidth\columnwidth
  \global\displaywidth\columnwidth}
\begin{document}
\draft 
\tighten

\title{Spin-polarized transport through a quantum dot coupled to ferromagnetic leads:
Kondo correlation effect}
\author{Jing Ma, Bing Dong, and X. L. Lei}
\address{Department of Physics, Shanghai Jiaotong University,
1954 Huashan Road, Shanghai 200030, China}
\date{\today}
\maketitle
\begin{abstract}
We investigate the linear and nonlinear transport through a single level quantum dot
connected to two ferromagnetic leads in Kondo regime, using the slave-boson
mean field approach for finite on-site Coulomb repulsion. We find that
for antiparallel alignment of the spin orientations in the leads, a
single zero-bias Kondo peak always appears in the voltage-dependent differential
conductance with peak height going down to zero as the polarization grows to $P=1$.
For parallel configuration, with increasing polarization from zero,
the Kondo peak descends and greatly widens by the appearance of shoulders,
and finally splits into two peaks on both sides of the bias voltage 
around $P\sim 0.7$ until disappears at
even larger polarization strength. At any spin orientation angle $\theta$, 
the linear conductance generally drops with growing polarization strength.
For a given finite polarization, the minimum linear conductance always 
appears at $\theta=\pi$.
\end{abstract}

\pacs{72.15.Qm, 73.23.Hk, 73.40.Gk, 73.50.Fq}

\narrowtext 
\section{Introduction}
Since the discovery of the giant magnetoresistance effect,\cite{GMR} 
extensive theoretical and experimental attention has been paid to 
the spin-dependent electron transport through systems consisting of 
two ferromagnetic layers (FM) sandwiched by a nonmagnetic structure 
of different types, including insulator,\cite{Jull} semiconductor quantum 
well,\cite{Wang} carbon nanotube,\cite{Tsuk} and composite structure.\cite{Sheng} 
Quite a lot interesting spin related transport effects and device functions have been 
predicted, observed, even realized. These investigations constitute an important part 
of the rapidly developing field of magnetoelectronics or spin-electronics.\cite{Prinz} 
Recently, FM/quantum-dot(QD)/FM system has attracted much attention. 
Different from previously studied nonmagnetic structure, a quantum dot features 
the strong many-body correlation among electrons in it such that, 
not only the Coulomb blockade can dominate its electronic transport,
but, particularly, a significant Kondo effect may arise when it connects 
to normal leads,
exhibiting zero-bias maximum in the differential conductance in the cases when  
odd number of electrons resides in. 
It is of course interesting to see what happens when the normal
metal leads are replaced by ferromagnetic leads. How is the Kondo-correlated state 
affected by the strength and the relative orientation of the spin polarizations of 
two magnetic leads?

In a recent paper, Sergueev {\it et al.}\cite{Serg} presented a theoretical
analysis of the transport characteristics of such a FM/QD/FM system modeled
by the Anderson single-impurity Hamiltonian with finite Coulomb repulsion $U$, 
using the ansatz proposed by Ng\cite{Ng} and the standard equation-of-motion 
technique for the retarded Green's function with the usual decoupling procedure 
for the higher order functions. They found that there is always a sharp single Kondo
resonant peak in nonlinear differential conductance at zero bias and
that the height of the peak exhibits only a modest change with varying 
the polarization strength from 0 to 0.6 and with varying the spin orientation angle
from 0 to $\pi$.
Very recently, markedly different results were reported on similar systems 
in the $U\rightarrow \infty$ limit by two groups,\cite{Mart,Lu} 
based also on the equation-of-motion technique together with Ng's ansatz\cite{Ng} 
or with some kind of additional assumption to determine renormalized level of the QD.
They found that for parallel alignment of the lead magnetizations, the Kondo resonance 
peak in the differential conductance splits at a polarization value as small as 0.2.
 
In this paper we investigate the spin-polarized transport through a quantum dot
coupled to ferromagnetic leads using a finite-$U$ slave-boson mean-field (SBMF)
approach of Kotliar and Ruckenstein.\cite{KR} As an effective nonperturbative 
technique, different versions of SBMF method\cite{Cole,KR} have been used
to study the equilibrium and out-of-equilibrium Kondo effect in a single QD 
and double QDs connected to normal leads without and with a magnetic 
field.\cite{Aono,Geo,Agu,Dong1,Dong2} 
It is generally believed that the SBMF approach is reliable in describing 
the Kondo regime at low temperature but may have problem 
to deal with strong charge fluctuations.\cite{Agu} 
Nevertheless,  previous analyses\cite{Dong1} on a single quantum dot with 
normal leads based on the finite-$U$ SBMF approach predicted the linear conductance 
in reasonable agreement with experiments\cite{Gold} and 
with numerical renormalization group calculation 
within the range $-1.2U \leq \epsilon_d \leq 0.2U$ of the dot level 
$\epsilon_d$, and predicted
the magnetic-field-induced peak splitting of nonlinear differential conductance
in Kondo regime in qualitatively agreement with experiments\cite{Gold} and with exact 
Bethe-ansatz solution\cite{Moor} in the voltage range up to several multiples of the 
Kondo temperature $T_K$. 
Even for shot noise the SBMF treatment was shown to yield good result
up to the bias voltage $eV\sim 0.8 T_K$.\cite{Meir2} 
This indicates that, although the SBMF approach should generally restrict to 
low temperatures and low voltages ($eV, T\leq T_K$),\cite{Meir2} 
the finite-$U$ SBMF method 
can be used to investigate the linear conductance within a relatively wide dot-level 
range and to study the nonlinear conductance for Kondo systems covering     
the bias range $eV$ up to several multiples of $T_K$, in which the main magnetic-field 
and magnetization induced features in Kondo transport appear.

Using the SBMF approach of Kotliar and Ruckenstein\cite{KR} we have investigated
the linear and nonlinear dc transport in FM/QD/FM systems having finite on-site 
Coulomb repulsion $U$. We find that the Kondo effect shows up as a broad zero-bias 
resonance in the nonlinear differential conductance versus bias voltage for antiparallel 
spin alignment in the leads. In parallel configuration, with increasing polarization
strength this zero-bias Kondo peak descends and greatly widens by the appearance of 
shoulders, and finally splits into two peaks on both sides of bias voltage 
at polarization strength around 0.7 until it disappears at even larger polarization. 
The linear conductance always drops with enhancing the 
polarization strength at any spin orientation angle,
and exhibits a strong angle variation at large polarization.
For a given finite value of polarization, 
the minimum linear conductance always appears at antiparallel configuration. 

\section{Formulation}

We consider a FM/QD/FM system similar to that discussed by Wang {\it et al.}\cite{Wang}: 
a quantum dot connected to two ferromagnetic leads. When a voltage $V$ is applied 
across two leads a current flows through the QD along the $x$ direction. 
The magnetic moment of the left lead is pointing to the $z$ direction, ${\bf 
M}_L=(0,0,M)$, 
while that of the right lead is at an angle $\theta$ to the $z$ axis in the $y$-$z$ 
plane,
${\bf M}_R=(0,M\sin\theta,M\cos\theta)$. The Hamiltonian of the system can be written 
as\cite{Wang}
\begin{equation}
H=H_L+H_R+H_D+H_T.
\end{equation}
Here 
\begin{equation}
H_{D}=\sum\limits_{\sigma}\epsilon_d c^{\dag}_{d\sigma}c_{d\sigma}+
U c^{\dag}_{d\uparrow}c_{d\uparrow}c^{\dag}_{d\downarrow}c_{d\downarrow}
\end{equation}
describes the QD with a single orbital level $\epsilon_d$ and a finite on-site Coulomb 
repulsion $U$ between electrons;
\begin{equation}
H_\alpha=\sum\limits_{k,\sigma}\epsilon_{\alpha k\sigma} C^{\dag}_{\alpha k\sigma}
C_{\alpha k\sigma}
\end{equation}
stands for the left ($\alpha=L$) or the right ($\alpha=R$) lead 
with electron energy $\epsilon_{\alpha k\sigma}=\epsilon_{\alpha k}+\sigma M$; 
and ($\bar{\sigma}=-\sigma$)
\begin{eqnarray}
H_T=\sum\limits_{k,\sigma}&&[
T_{{R}k\sigma}(\cos{\frac{\theta}{2}}C^{\dag}_{Rk\sigma}-\sigma\sin{\frac{\theta}{2}}
C^{\dag}_{Rk{\bar{\sigma}}})c_{d\sigma}\nonumber\\
&+&T_{Lk\sigma}C^{\dag}_{Lk\sigma}c_{d\sigma}+\,h.c.]
\end{eqnarray}
describes the tunneling between the QD and the two leads.
Note that the small letter symbols $c^{\dag}_{d\sigma}$ ($c_{d\sigma}$) and 
$c^{\dag}_{\alpha k\sigma}$ ($c_{\alpha k\sigma}$) are creation
(annihilation) operators of electrons in the dot and in the left  
and right leads with spin up ($\sigma=1$ or $\uparrow$) or 
spin down ($\sigma=-1$ or $\downarrow$) state in respect to the $z$-axis.
In the above expressions, the capital letter operator $C^{\dag}_{Lk\sigma}=
c^{\dag}_{Lk\sigma}$ for the left lead, while $C^{\dag}_{Rk\sigma}=
\cos(\theta/2)c^{\dag}_{Rk\sigma}+\sigma\sin(\theta/2)c^{\dag}_{Rk\bar{\sigma}}$
for the right lead.

According to the finite-$U$ slave-boson approach,\cite{KR},
one can use additional four auxiliary boson operators $e$, $p_\sigma \,(\sigma=\pm1)$ 
and $d$, which are associated respectively with the empty, singly occupied, and doubly 
occupied electron states of the QD, to describe the above physical problem without 
interparticle coupling in an enlarged space with constraints: the completeness relation 
$\sum\limits_\sigma p^\dag_{\sigma}p_\sigma+e^{\dag}e+d^{\dag}d=1$  
and the particle number conservation condition 
$c^{\dag}_{d\sigma}c_{d\sigma}=p^{\dag}_{\sigma}p_\sigma+d^{\dag}d,\,(\sigma=\pm1)$.
Within the mean-field scheme we start with the following effective Hamiltonian:
\widetext
\begin{eqnarray}
H_{eff}&=&\sum\limits_{\sigma}\epsilon_{d}c^{\dag}_{d\sigma}c_{d\sigma}+
Ud^{\dag}d +\sum\limits_{\alpha=L,R}H_{\alpha}
+\lambda^{(1)}(\sum\limits_{\sigma}p^\dag_{\sigma}p_\sigma+e^{\dag}e+d^{\dag}d-1)
 +\sum\limits_{\sigma}\lambda^{(2)}_{\sigma}(c^{\dag}_{d\sigma}c_{d\sigma}-
 p^{\dag}_{\sigma}p_\sigma-d^{\dag}d)\nonumber\\
&+&\sum\limits_{k,\sigma}[T_{Lk\sigma}C^{\dag}_{Lk\sigma}
c_{d\sigma}z_\sigma+T_{{R}k\sigma}(\cos{\frac{\theta}{2}}C^{\dag}_{Rk\sigma}
 -\sigma\sin{\frac{\theta}{2}}C^{\dag}_{Rk\bar{\sigma}})c_{d\sigma}z_\sigma+\,h.c.]
\end{eqnarray}
\narrowtext
where three Lagrange multipliers $\lambda^{(1)}$ and $\lambda^{(2)}_\sigma$ are 
introduced to take account of the constraints, and in the hopping term the 
QD fermion operators $c^{\dag}_{d\sigma}$ and $c_{d\sigma}$ are expressed as 
$z_\sigma^{\dag}c_{d\sigma}^{\dag}$ and $c_{d\sigma}z_\sigma$ to recover the 
many body effect on tunneling. $z_\sigma$ consists of all boson operator sets 
that are associated with the physical process with which a $\sigma$-spin electron
is annihilated: 
\begin{equation}
z_\sigma=(1-d^{\dag}d-p^{\dag}_{\sigma}p_{\sigma})^{-\frac{1}{2}}(e^{\dag}p_\sigma+
p^{\dag}_{\bar{\sigma}}d)(1-e^{\dag}e-p^{\dag}_{\bar{\sigma}}p_{\bar{\sigma}})
^{-\frac{1}{2}}.
\end{equation}

From the effective Hamiltonian (5) one can derive four equations of motion of 
slave-boson operators, which, together with the three constraints, serve as the basic
equations. Then we use the mean-field approximation in the statistical 
expectations of these equations, in which all the boson operators are replaced 
by their expectation values. In the wide-band limit for the leads 
the resulting equations are as follows.
\begin{eqnarray}
\sum\limits_\sigma\frac{{\partial}z_\sigma}{{\partial}e}K_\sigma+2\lambda^{(1)}e=0,&&\\
\sum\limits_\sigma\frac{{\partial}z_\sigma}{{\partial}p_{\sigma'}}K_\sigma+
2(\lambda^{(1)}-\lambda^{(2)}_{\sigma'})p_{\sigma'}=0,&&\,\,\sigma'=\pm1,\\
\sum\limits_\sigma\frac{{\partial}z_\sigma}{{\partial}d}K_\sigma+2(U+\lambda^{(1)}-
\sum_\sigma\lambda^{(2)}_\sigma)&&d=0,\\
\sum\limits_{\sigma}|p_\sigma|^2+|e|^2+|d|^2=1,&&\\
\frac{1}{2{\pi}i}\int{d\omega}G^{<}_{d\sigma\sigma}(\omega)=|p_\sigma|^2+|d|^2,&&\,\,
 \sigma=\pm1.
\end{eqnarray}
Here
\widetext
\begin{equation}
K_\sigma \equiv \frac{1}{\pi} \int d\omega \Big( z_\sigma{\rm Re}(G^{r}_{d\sigma\sigma})
\big[f_R (\Gamma_{R\sigma}\cos^2{\frac{\theta}{2}}+
\Gamma_{R{\bar{\sigma}}}\sin^2{\frac{\theta}{2}})+
 f_L \Gamma_{L\sigma}\big] 
  + \sigma z_{\bar{\sigma}} {\rm Re}(G^{r}_{d\sigma{\bar{\sigma}}}) f_R  
  (\Gamma_{R\sigma}-\Gamma_{R{\bar{\sigma}}})
  \sin{\frac{\theta}{2}}\cos{\frac{\theta}{2}}\Big), \nonumber
\end{equation}
\narrowtext
\noindent
with $f_\alpha(\omega)=1/(e^{\beta(\omega-\mu_\alpha)}+1)$, $\mu_\alpha$
the chemical potential of the $\alpha$th lead, which is assumed in an
equilibrium state at temperature $1/\beta$, and 
$\Gamma_{\alpha\sigma}(\omega)=
2\pi\sum_{k\in\alpha}|T_{\alpha k\sigma}|^2\delta(\omega-\epsilon_{\alpha k\sigma})$
the coupling strength function between the QD and the lead $\alpha$.
$G^{r(a)}_{d\sigma\sigma'}(\omega)$ 
and $G^<_{d\sigma\sigma'}(\omega)$ are the elements of the $2\times 2$
retarded (advanced) and correlation Green's function matrices
${\bf G}^{r(a)}_d (\omega)$ and ${\bf G}^{<}_d (\omega)$ in the spin
space of the QD.  The retarded (advanced) Green's function can
be written in the form renormalized due to dot-lead couplings,
\begin{equation}
{\bf G}^{r(a)}_d (\omega)=\frac{1}{\omega{\bf I}-\tilde{\bf H}_{d}\pm i\tilde{\bf 
\Gamma}}
\end{equation}
in which ${\bf I}$ is a unit matrix,
\begin{equation}
\tilde{\bf H}_{d}=
\left(
\begin{array}{cc}
\tilde{\epsilon}_{d\uparrow}&0 \\ 0&\tilde{\epsilon}_{d\downarrow}
\end{array}
\right)
\end{equation}
 reflects the dot-level shifting and splitting between spin-up and spin-down
 electrons,
$\tilde{\epsilon}_{d\sigma}=\epsilon_{d}+\lambda^{(2)}_\sigma$, and 
$\tilde{\bf \Gamma}=\frac{1}{2}(\tilde{\bf \Gamma}_L+\tilde{\bf \Gamma}_R)$
is the effective line width or the renormalized co-tunneling strength,
where $\tilde{\bf \Gamma}_\alpha$  ($\alpha=L$ or $R$) is a $2\times 2$ matrix 
consisting of elements $\tilde{\bf \Gamma}_\alpha^{11}=
|z_\uparrow|^2(\Gamma_{\alpha\uparrow}\cos^2{\frac{\theta_\alpha}{2}}+
\Gamma_{\alpha\downarrow}\sin^2{\frac{\theta_\alpha}{2}})$,
$\tilde{\bf \Gamma}_\alpha^{22}=
|z_\downarrow|^2(\Gamma_{\alpha\downarrow}\cos^2{\frac{\theta_\alpha}{2}}+
\Gamma_{\alpha\uparrow}\sin^2{\frac{\theta_\alpha}{2}})$, and
$\tilde{\bf \Gamma}_\alpha^{12}=\tilde{\bf \Gamma}_\alpha^{21}=
z_{\uparrow}z_{\downarrow}(\Gamma_{\alpha\uparrow}-\Gamma_{\alpha\downarrow})
\cos{\frac{\theta_\alpha}{2}}\sin{\frac{\theta_\alpha}{2}}$, where
$\theta_L=0$ and $\theta_R=\theta$.
The correlation Green's function ${\bf G}^{<}_d$ can be obtained
with the help of the following relation typical for a noninteracting system:
\begin{equation}
{\bf G}_d^{<}={\bf G}_d^r {\bf \Sigma}_{d}^{<}{\bf G}_d^a,
\end{equation}
with ${\bf \Sigma}_d^{<}=i(\tilde{\bf \Gamma}_L f_L+\tilde{\bf \Gamma}_R f_R)$.

The electric current flowing from the left lead into the QD can be obtained
from the rate of change of electron number operator of the left lead:\cite{Meir,Jauho}
\widetext
\begin{equation}
I_L=\frac{ie}{\hbar}\int\frac{d\omega}{2\pi}{\rm Tr}\Big\{ 
{\tilde{\bf \Gamma}}_L(\omega)
\big( [ {\bf G}^r_d(\omega)-{\bf G}^a_d(\omega)]f_L(\omega)+
{\bf G}^{<}_d(\omega)\big)\Big\}.
\end{equation}
\narrowtext
In the steady transport state, the current flowing from the QD to the right lead must 
be equal to the current from the left lead to the QD, and the formula (17) can be 
directly used for calculating the current flowing through the lead-dot-lead system 
under a bias voltage $V$ between the two leads:
$I=I_L$.

\section{Numerical results and discussions}
  We suppose that the left and right leads are made from the identical material and, 
in the wide band limit, assume constant effective coupling strength
$\Gamma_{L\sigma}(\omega)=\Gamma_{R\sigma}(\omega)=\Gamma_{\sigma}$ respectively
for up- and down-spin orientation. 
The steady state dc current flowing through the system reads 
\begin{equation}
I=-\frac{e}{h}{\int}d\omega\sum_{\sigma} \Gamma_\sigma|z_\sigma|^2[
2 f_L{\rm Im}(G^r_{d\sigma\sigma})+{\rm Im}(G^{<}_{d\sigma\sigma})].
\end{equation} 
We will take $\Gamma\equiv(\Gamma_\uparrow+\Gamma_\downarrow)/2$
as the energy units in the following and define the spin polarization as
$P\equiv (\Gamma_\uparrow-\Gamma_\downarrow)/(2 \Gamma)$. The Kondo temperature
in the case of $P=0$, given by
 $T_K^0=U\sqrt{\beta} \exp(-\pi/\beta)/2\pi$ with 
$\beta=-2U\Gamma/\epsilon_d(U+\epsilon_d)$,
will be used as the dynamical energy scale in presenting the nonlinear conductance.
 
 In the following we will deal with FM/QD/FM systems having a fixed finite Coulomb
repulsion $U=6$ and consider effects of changing bare dot level $\epsilon_d$, 
polarization strength $P$ and relative spin orientation $\theta$.   

In calculating the dc current from Eq.\,(17) under a finite bias voltage $V$ between 
the two leads, we choose the energy zero such that the chemical potential 
$\mu_L=-\mu_R=eV/2$ for the left and right leads for convenience.

Linear conductance $G=(dI/dV)_{V=0}$ is related to the slave-boson parameters 
$e^2,p_\sigma^2, d^2, \lambda^{(1)}$, $\lambda^{(2)}_\sigma$ and $|z_\sigma|^2$ 
at zero bias. In antiparallel configuration ($\theta=\pi$), all the slave-boson 
parameters are identical for up and down spin indices and independent of $P$. 
In the parallel configuration ($\theta=0$), $p_\sigma^2, \lambda^{(2)}_\sigma$ 
and $|z_\sigma|^2$ split for up and down spins and all the parameters exhibit 
strong $P$-dependence. As an example we plot in Fig.1 the zero-bias parameters 
$e^2,p_\sigma^2, d^2, \lambda^{(1)}$, $\lambda^{(2)}_\sigma$ and the average electron 
number in the dot $n=p_\uparrow^2+p_\downarrow^2+2d^2$ in the parallel configuration 
at zero temperature for the case of $\epsilon_d=-1$ and $U=6$. 
\begin{figure}[htb]
\hspace*{0.8cm}
  \psfig{figure=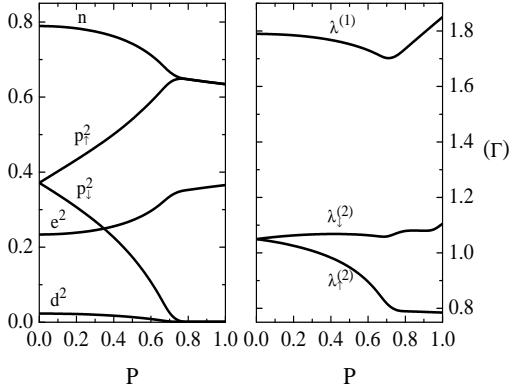,width=7.0cm,height=5.5cm,angle=0,clip=on}
  \vspace{0.2cm}
    \caption{The expectation values of slave-boson operators
$e^2, d^2, p_{\uparrow}^2, p_{\downarrow}^2$, the corrections of the
renormalized dot level $\lambda^{(1)},\lambda^{(2)}_\uparrow,\lambda^{(2)}_\downarrow$, 
and the electron number dwelling in the dot $n=p_\uparrow^2+p_\downarrow^2+2d^2$, are 
shown as functions of polarization $P$ 
at zero temperature and zero bias voltage $V=0$. 
The quantum-dot in the FM/QD/FM system has a single bare energy level $\epsilon_d=-1$ 
and a 
finite on-site Coulomb repulsion $U=6$.}
\end{figure}

Figs.\,2 and 3 show the zero-temperature linear conductance $G$ as a function of the 
bare
energy level $\epsilon_d$ of the quantum-dot having a fixed on-site Coulomb repulsion
$U=6$ at different polarizations $P=0,0.3,0.5,0.7$ and $0.9$ on two leads
in parallel ($\theta=0$) and in antiparallel ($\theta=\pi$) configurations. 
Each curve covers three regimes, containing the resonance peak due to the dot level 
around $\epsilon_d=0$, the charging peak around $\epsilon_d=-U$, and the Kondo 
peak centered at $\epsilon_d=-U/2$. 
In the case of zero polarization $P=0$ and in the antiparallel configuration of
 arbitrary polarization, all the slave-boson parameters are identical for up 
 and down spin indices, and electrons with up-spin and down-spin are equally 
 available in the whole lead-dot-lead system, favoring the formation of the 
 Kondo-correlated state within a relatively wide dot level range centered 
 at $\epsilon_d=-3$. At the same time, 
since there is no splitting of the renormalized dot levels $\tilde{\epsilon}_{d\sigma}$ 
for up and down spins, the resonance and charging peaks are relatively narrow. 
The $G$-vs-$\epsilon_d$ curves appear to be smooth hump-type structures.  
On the other hand, since there is no spin-flip mechanism in the tunneling coupling 
and in the antiparallel configuration the available minority-spin (e.g. up-spin) 
states in the right lead decrease with increasing polarization strength, 
the transfer of the majority-spin (up-spin) electrons from the left lead to 
the right lead is suppressed by the finite $P$, such that the conductance 
of the system goes down with increasing $P$ and vanishes at $P=1$. 
In the parallel configuration, finite polarization $P>0$ splits the dot level 
for up and down spins and thus broadens the resonance peaks around $\epsilon_d=0$ 
and $\epsilon_d=-U$. At the same time the central Kondo peaks is progressively
narrowed with increasing $P$ due to the decrease of the available minority-spin
electrons in the two leads. This two factors 
lead to the appearance of kinks or splitting peaks in the $G$-vs-$\epsilon_d$
curves for $P=0.3, 0.5, 0.7$ and $0.9$ in the parallel configuration. Nevertheless,
the unitary limit $G=2\,e^2/h$ can still be reached as long as $P<1$. 
In the case of $P=1$, since there is no minority-spin electron in the leads, 
the formation of the Kondo-correlated state is impossible and the double occupancy 
probability of the dot level vanishes, $d^2=0$. In this case 
the Kondo peak disappears, together with the charging peak. There remains
only a tunneling peak centered around $\epsilon_d=0$ in the $G$-vs-$\epsilon_d$
curve, which stems from the up-spin electrons.
\begin{figure}[htb]
\hspace*{0.8cm}
  \psfig{figure=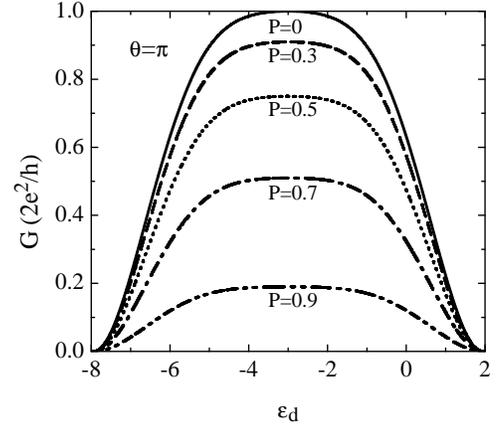,width=6.7cm,height=6.0cm,angle=0,clip=on}
  \vspace{0.2cm}
    \caption{The linear conductance $G$ of FM/QD/FM systems is shown as a function 
	of the bare dot level $\epsilon_d$ for different spin polarization $P$ of the
	 leads in the antiparallel configuration ($\theta=\pi$). The dot has a finite
	  on-site Coulomb repulsion $U=6$.}
\end{figure}
\begin{figure}[htb]
\hspace*{0.8cm}
  \psfig{figure=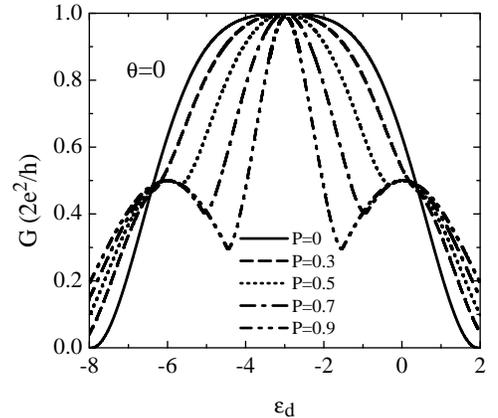,width=6.7cm,height=6.0cm,angle=0,clip=on}
  \vspace{0.2cm}
    \caption{The linear conductance $G$ of FM/QD/FM systems is shown as a function 
	of the bare dot level $\epsilon_d$ for different spin polarization $P$ of the
	 leads in the parallel configuration ($\theta=0$). The dot has a finite on-site 
Coulomb repulsion $U=6$.}
\end{figure}
Fig.\,4 shows the calculated zero-temperature linear conductance $G$ for a FM/QD/FM
system having $\epsilon_d=-2$ and $U=6$
as a function of polarization $P$ at several different spin orientations 
$\theta=0,\pi/4,\pi/2,3\pi/4$, and $\pi$. For each fixed orientation  
the linear conductance decreases with increasing $P$ straightforwardly from the 
maximum value $G=1.97\,e^2/h$ 
at $P=0$ down to its minimum. The conductance descends 
generally faster at larger $\theta$ except for the case of 
$\theta=\pi/4$ and $P>0.75$, where $G$ goes down slower than that of $\theta=0$.
This can be understood as due to the rapid decrease of the conductance (or the 
swift narrowing of the Kondo peak) with $P$ increasing in the parallel configuration, 
as seen in Fig.\,3. 
For a given large $P$ when rotating an angle $\pi/4$ from parallel configuration, 
the conductance increase due to the reduction of the equivalent parallel 
polarization from $P$ to $P\cos(\pi/4)$ overcompensates the conductance decrease 
induced by applying an equivalent polarization $P\sin(\pi/4)$ 
along $\theta=\pi/2$.
In Fig.\,5 we plot the linear conductance $G$ versus the angle $\theta$ of the
relative spin orientation at various polarization strengths $P=0,0.3,0.5,0.7$ and $0.9$
for the same system as described in Fig.\,4.
We see a clear angle variation of the linear conductance at a finite spin polarization,
with the minimum always at $\theta=\pi$. At small $P$ the conductance $G$ reaches 
its maximum at $\theta=0$, but at large polarization the maximum appears near
 $\theta=\pi/4$.
\begin{figure}[htb]
\hspace{0.3cm}
  \psfig{figure=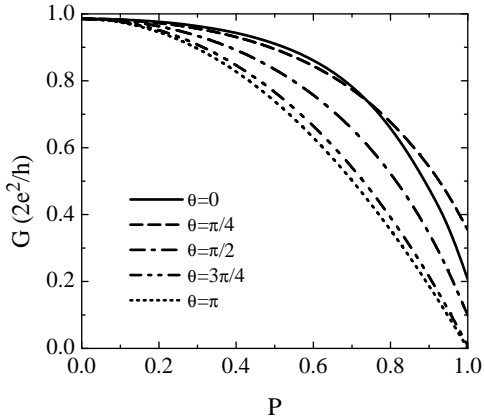,width=6.7cm,height=6.0cm,angle=0,clip=on}
   \caption{The linear conductance $G$ is shown as a function of
polarization $P$ for different spin orientation $\theta$ at zero temperature. 
The system has a single bare dot level $\epsilon_d=-2$ and a finite on-site Coulomb 
repulsion $U=6$.}
\end{figure}

Fig.\,6 shows the calculated differential conductance $dI/dV$ 
versus the bias voltage under antiparallel configuration
($\theta=\pi$) at various polarizations $P=0,0.3,0.5,0.7$ and $0.9$
for a FM/QD/FM system having a single dot level $\epsilon_d=-1$ and a finite
on-site Coulomb repulsion $U=6$ ($T_K^0=0.4$). 
All the curves exhibit a single zero-bias peak having essentially the same width 
but with peak height going straight down with growing $P$ from the highest value 
$1.79\,e^2/h$ at $P=0$.
As pointed above, in antiparallel configuration 
electrons with up-spin and down-spin are equally 
available in the lead-dot-lead system, in favor of the formation of the Kondo-correlated 
state for all values of $P$. However, since the up-spin states are almost unavailable 
in the right lead in the case of large polarization $P$, the transfer of electrons
(almost up-spins) from the left lead to the right lead is largely suppressed. 
The conductance of the system goes down to zero with increasing $P$ to 1.
\begin{figure}[htb]
\hspace{0.2cm}
  \psfig{figure=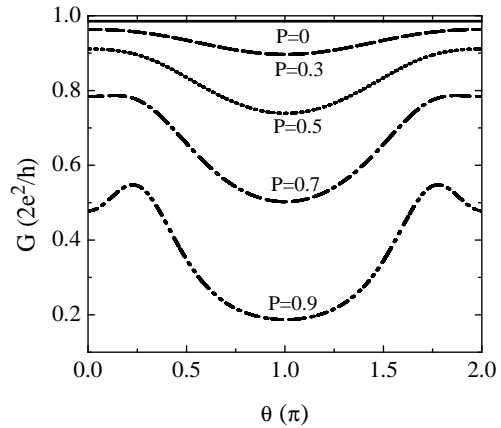,width=6.7cm,height=6.0cm,angle=0,clip=on}
   \caption{The linear conductance $G$ versus the relative spin polarization 
   angle $\theta$ at different polarization strength $P$ for the same system 
   as described in Fig.\,4.}
\end{figure}      
\begin{figure}[htb]
\hspace{0.3cm}
  \psfig{figure=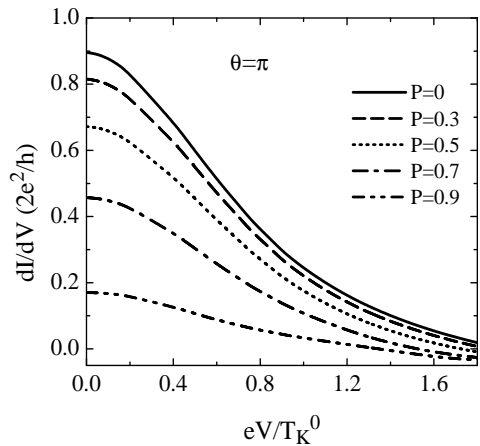,width=6.7cm,height=6.3cm,angle=0,clip=on}
   \caption{Zero-temperature differential conductance $dI/dV$ versus the bias 
   voltage $eV$ 
at different polarization $P$ in the antiparallel spin configuration ($\theta=\pi$). 
The quantum-dot in the FM/QD/FM system has a single bare energy level $\epsilon_d=-1$ 
and a finite on-site Coulomb repulsion $U=6$.}
\end{figure}

The situation becomes different in the case of parallel configuration, 
where the available down-spin electrons are less than the up-spin ones in the leads
when $P>0$.  
Although the former can still provide screening for dot electrons to form 
Kondo correlation to certain degree at small values of $P$, the effect gradually 
weakens with growing $P$ for this $\epsilon_d=-1$ system (somewhat away 
from the unitary limit),  
leading to the decrease in linear conductance from the peak value at $P=0$.
In the case of large polarization, the number of minority-spin electrons
is too small to form the Kondo-correlation state and Kondo-induced 
conductance enhancement disappears rapidly with $P$ increasing. 
On the other hand, even when there is no down-spin states ($P=1$), the up-spin states 
are always available in both left and right leads, allowing the up-spin electrons to
carry a current through tunneling.  
Fig.\,7 illustrates the calculated differential conductance $dI/dV$
for $\theta=0$ at various polarizations $P=0,0.3,0.5,0.7$ and $0.9$. 
We can see that the zero-bias Kondo peak in the $dI/dV$-vs-$eV$ curve of $P=0$,
descends and widens by the appearance of shoulders on both sides of bias voltage 
at $P=0.3$ and $P=0.5$. This is apparently due to the weakening of the Kondo
correlation and the splitting of the renormalized dot level $\tilde{\epsilon}_d$ 
(see Fig.\,1).  
At $P=0.7$, the level splitting is large enough and there are still sufficient
down-spin electrons for the formation of Kondo state, we see the splitting of the
zero-bias anomaly into two peaks on both sides of the applied voltage 
at a distance around twice the renormalized level splitting 
($\lambda_\downarrow^{(2)}-\lambda_\uparrow^{(2)}\approx 0.27$) (see Fig.\,1).
At $P=0.8$, almost no down-spin electron exists in the dot and no
Kondo-related $dI/dV$ behavior shows up for this $\epsilon_d=-1$ system.
\begin{figure}[htb]
\hspace{0.3cm}
  \psfig{figure=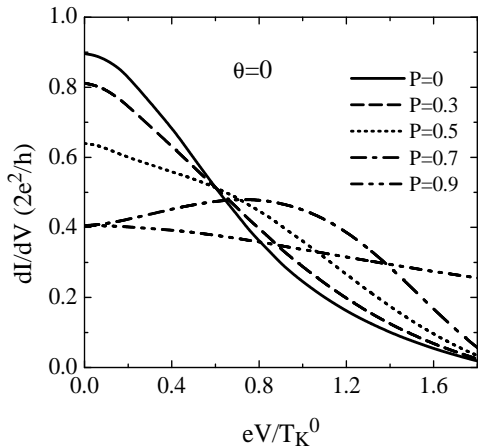,width=6.7cm,height=6.3cm,angle=0,clip=on}
   \caption{Zero-temperature differential conductance $dI/dV$ versus the bias 
   voltage $eV$ 
at different polarization $P$ in the parallel spin configuration ($\theta=0$). 
The quantum-dot in the FM/QD/FM system has a single bare energy level $\epsilon_d=-1$ 
and a finite on-site Coulomb repulsion $U=6$.}
\end{figure}

Similar trend can also be seen in Fig.\,8, where we plot the differential
conductance $dI/dV$ versus the bias voltage $eV$ in the case of $P=0.8$ at various
relative spin orientations $\theta=0,\pi/4,\pi/2,3\pi/4$ and $\pi$. 
At this strength of polarization, there is no Kondo-correlated state to appear for
the $\epsilon_d=-1$ system in the parallel configuration ($\theta=0$).  
In the case of $\theta=\pi/4$, however, the dot level splitting is
large enough and there are sufficient down-spin electrons to survive the reduced
effective parallel polarization $P\cos(\pi/4)$ for the formation of the Kondo
correlation. A clear appearance of nonzero-bias maximum in the voltage-dependent
 conductance can be seen. 
\begin{figure}[htb]
\hspace{0.3cm}
  \psfig{figure=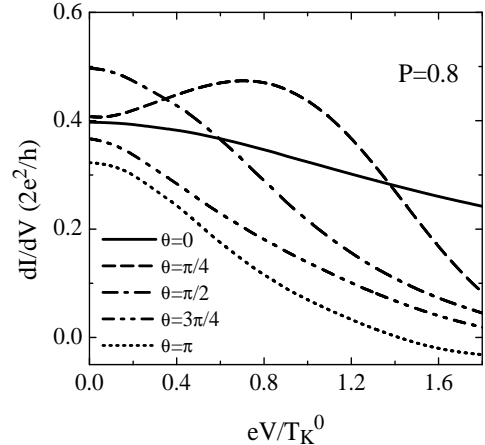,width=6.7cm,height=6.3cm,angle=0,clip=on}
   \caption{Zero-temperature differential conductance $dI/dV$ versus the bias 
   voltage $eV$ 
at different spin orientation $\theta$ for polarization strength $P=0.8$. 
The quantum-dot in the FM/QD/FM system has a single bare energy level $\epsilon_d=-1$ 
and a finite on-site Coulomb repulsion $U=6$.}
\end{figure}
 
In spite of dealing with the same system ($U=6$ and $\epsilon=-1$), the behavior 
of the linear and nonlinear conductance obtained here is quite 
different from that reported in Ref.\onlinecite{Serg}.
In parallel configuration, our predicted linear conductance $G$ goes down with
increasing $P$ at least up to $P=0.7$ (see Fig.\,7) in contrast to their $G$ which
increases with rising $P$ (Fig.\,5a in Ref.\onlinecite{Serg}). In the bias-dependent 
differential conductance our results show the clear trend of the zero-bias peak 
broadening with increasing $P$ and finally splitting at large $P$ under parallel and 
nearly parallel spin alignments due to the splitting of the renormalized dot level,
while Ref.\,\onlinecite{Serg} predicts a single zero-bias peak narrowing with growing 
$P$. In antiparallel configurations we obtain a much wider zero-bias peak in nonlinear
differential conductance than theirs. Our predicted splitting of the zero-bias anomaly
in nonlinear differential conductance under parallel and nearly parallel spin alignments 
for large $P$, is in general agreement with Refs.\,\onlinecite{Mart,Lu} in the large $U$ 
limit. However, in both parallel and antiparallel configurations,
the zero-bias peaks and the split peaks seen in the voltage-dependent differential 
conductance in the present investigation are broader than those reported in 
Refs.\,\onlinecite{Mart,Lu}. It should be noted that the present analysis yields
a smaller dot-level spltting than theirs. For instance, for systems with 
$U=\infty$, $\epsilon_d=-2$, Ref.\onlinecite{Mart} reported a distance $e\Delta V$ 
around $0.5$ between the split peaks in the voltage-dependent differential conductance 
in the parallel configuration at a polarization strength of $P=0.2$, that is about 
the same  split-peak distance obtained at $P=0.6$ in this paper for the system 
of $U=6$ and $\epsilon_d=-1$.   

\section{Summary}
We have theoretically investigated the linear and nonlinear electron transport 
of a spin-valve system consisting of a quantum dot connected to two
ferromagnetic leads in Kondo regime but somewhat apart from the unitary
limit. Based on the finite-$U$ slave-boson mean field approach we find
markedly different transport behavior when changing the relative spin
orientation. In the antiparallel configuration where
electrons with up-spin and down-spin are equally available in the system,
a single zero-bias Kondo peak always appears in the voltage-dependent differential
conductance through the whole range of polarization $0\leq P < 1$, but 
the peak height drops down with increasing $P$ and vanishes at $P=1$.
In the parallel configuration, with increasing the spin polarization
from zero, Kondo peak descends and greatly widens by the appearence
of shoulders and finally splits into two peaks on both sides of bias voltage
at polarization around $P\sim 0.7-0.8$, 
until disappears at even larger $P$.
At any spin alignment angle, the linear
conductance generally drops with increasing spin polarization.
It is shown that a Kondo-dominant FM/QD/FM system may exhibit
strong angle variation in linear conductance at large polarization,
forming an effective spin valve.

We would like to thank Dr. S.Y. Liu and Dr. W.S. Liu for helpful discussions.
This work was supported by the National Science Foundation of China,
the Special Funds for Major State Basic Research Project, and
the Shanghai Municipal Commission of Science and Technology.

\end{multicols}
\end{document}